\documentclass[aps,pra,superscriptaddress,showpacs,twocolumn]{revtex4-1}
\usepackage{amssymb}
\usepackage{graphicx}
\usepackage{dcolumn}
\usepackage{bm}
\usepackage{amsmath}

\usepackage[usenames]{color}
\usepackage{textcomp}
\usepackage{float}
\def\be{\begin{equation}}
\def\ee{\end{equation}}
\def\bea{\begin{eqnarray}}
\def\eea{\end{eqnarray}}

\usepackage[unicode=true,bookmarks=true,bookmarksnumbered=false,bookmarksopen=false,breaklinks=false,pdfborder={0 0 1},backref=false,colorlinks=true]{hyperref}

\hypersetup{linkcolor=magenta,urlcolor=blue,citecolor=blue,pdfstartview={FitH},hyperfootnotes=false,unicode=true}

\begin{document}
\title{Quantum phases of spin-orbital-angular-momentum coupled bosonic gases in optical lattices}
\author{Rui Cao}
\affiliation{Department of Physics, National University of Defense Technology, Changsha 410073, P. R. China}
\author{Jinsen Han}
\affiliation{Department of Physics, National University of Defense Technology, Changsha 410073, P. R. China}
\author{Jianhua Wu}
\email{wujh@nudt.edu.cn}
\affiliation{Department of Physics, National University of Defense Technology, Changsha 410073, P. R. China}
\author{Jianmin Yuan}
\affiliation{Department of Physics, Graduate School of China Academy of Engineering Physics, Beijing 100193, P. R. China}
\affiliation{Department of Physics, National University of Defense Technology, Changsha 410073, P. R. China}
\author{Lianyi He}
\email{lianyi@mail.tsinghua.edu.cn}
\affiliation{Department of Physics and State Key Laboratory of Low-Dimensional Quantum Physics, Tsinghua University, Beijing 100084, China}
\author{Yongqiang Li}
\email{li\_yq@nudt.edu.cn}
\affiliation{Department of Physics, National University of Defense Technology, Changsha 410073, P. R. China}

\begin{abstract}
Spin-orbit coupling plays an important role in understanding exotic quantum phases. In this work, we present a scheme to combine spin-orbital-angular-momentum (SOAM) coupling and strong correlations in ultracold atomic gases.
Essential ingredients of this setting is the interplay of SOAM coupling and Raman-induced spin-flip hopping, engineered by lasers that couples different hyperfine spin states. In the presence of SOAM coupling only, we find rich quantum phases in the Mott-insulating regime, which support different types of spin defects such as spin vortex and composite vortex with antiferromagnetic core surrounded by the outer spin vortex. Based on an effective exchange model, we find that these competing spin textures are a result of the interplay of Dzyaloshinskii-Moriya and Heisenberg exchange interactions. In the presence of both SOAM coupling and Raman-induced spin-flip hopping, more many-body phases appear, including canted-antiferromagnetic and stripe phases. Our prediction suggests that SOAM coupling could induce rich exotic many-body phases in the strongly interacting regime.
%In Superfluid, there are normal Superfluid and vortex Superfluid with rotating phase of order parameter$\Phi$.
\end{abstract}

\date{\today}

%\pacs{67.85.-d, 03.75.Mn, 05.30.Jp, 05.30.Rt}

\maketitle
\section{Introduction}
Spin-orbit coupling, the interplay of particle's spin and orbital degrees of freedom, plays a crucial role in various exotic phenomena in solid-state systems, such as the quantum spin Hall effect~\cite{science.1133734,konig2007quantum,PhysRevLett.94.047204,science.1105514}, topological insulators~\cite{RevModPhys.82.3045}, and topological superconductors~\cite{RevModPhys.83.1057}. Ultracold atomic system, with high controllability degrees of freedom, is also a versatile candidate to investigate these quantum phenomena, by overcoming the problem of their neutrality~\cite{RevModPhys.83.1523}. One of these schemes relies on two-photon Raman transitions between two hyperfine states (pseudospin) of atoms~\cite{PhysRevA.79.063613}, which are coupled with the atomic center-of-mass momentum~\cite{Goldman_2014}. Here, propagation directions of laser beams are crucial to determine the type of spin-orbit coupling in ultracold atoms. When two beams counter-propagate, atom's spin can be coupled with linear momentum of atoms, i.e. spin-linear-momentum coupling~\cite{RevModPhys.83.1523,Goldman_2014,RevModPhys.91.015005,galitski2013spin}. Rich exotic quantum states have been observed in ultracold atomic gases with spin-linear-momentum coupling~\cite{lin2011spin,PhysRevLett.109.095301,PhysRevLett.109.095302,doi:10.1126/science.aaf6689,2016Experimental,li2017stripe}.

%The key technique for different SOC in ultracold atom is Raman coupling between different internal spin states with different momentum transferring to atoms as the spin state changes. In SLM coupling, atoms get photo momentum with the Raman process, the atom's spin and motional degrees of freedom become coupled.

Another fundamental type of spin-orbit coupling is called SOAM coupling. This coupling can be achieved by a pair of copropagating Laguerre-Gaussian (LG) lasers, where LG beam modes carry different orbital angular momenta along the direction of beam propagation~\cite{allen2016optical,Liu2006}. The atomic system obtains orbital angular momentum from the copropagating LG beams via Raman transitions among the internal hyperfine states of atoms, whereas the transfer of photon momentum into atoms is suppressed~\cite{PhysRevLett.121.113204,PhysRevLett.122.110402}. Within SOAM coupling, several intriguing quantum phases have been predicted theoretically~\cite{PhysRevA.93.013629,PhysRevA.94.033627,PhysRevA.91.063627,PhysRevResearch.2.033152,Chiu_2020,PhysRevA.91.033630,PhysRevA.92.033615,PhysRevA.102.063328,PhysRevA.102.013316,PhysRevLett.125.260407,PhysRevLett.126.193401,Bidasyuk_2022} and observed experimentally~\cite{PhysRevLett.121.113204,PhysRevLett.122.110402,PhysRevLett.121.250401,https://doi.org/10.48550/arxiv.2202.09017}. In these studies, however, interactions between atoms play tiny role in the various quantum phases, and one mainly focus on the weakly interacting regime.

In the paper, we combine SOAM coupling and strong correlations in ultracold gases, and focus on the response of spin degree of freedom to SOAM coupling. To achieve this goal, we propose a setup by introducing a beam with orbital angular momentum in the third direction ($z$ direction) for a two-component ultracold bosonic gas loaded into a blue-detuned square lattice, as shown in Fig.~\ref{1}. By controlling the frequency difference between the standing wave in the $x$ direction and the Raman beam in the $z$ direction, the two hyperfine states that match the Raman selection rules can be coupled, as shown in Fig~\ref{1}(b). In this setup, we actually achieve both a SOAM coupling in the $z$ direction~\cite{PhysRevLett.121.113204,PhysRevLett.122.110402}, and a Raman lattice in the $x$ direction~\cite{PhysRevLett.110.076401}. The competition between SOAM coupling and Raman-assisted spin-flip hopping may give rise to various quantum many-body phases.

This system can be effectively modeled by an extended Bose-Hubbard model for a sufficient deep optical lattice. We specifically consider the case of half filling in the Mott regime. To obtain many-body phases of the system, a bosonic version of real-space dynamical mean-field theory (RBDMFT) is implemented. Various competing phases are obtained in the Mott-insulating regime, including canted-antiferromagnetism, spin-vortex, and composite spin-vortex with {\it nonrotating} core. To explain the many-body phases, an effective spin-exchange model is derived, and we attribute these competing spin textures to the interplay of Heisenberg exchange and Dzyaloshinskii-Moriya interactions. Upon increasing the hopping amplitudes, atoms delocalize, and superfluid phases appear, including normal superfluid, rotating superfluid with vortex texture, and stripe superfluid.
%We have found that most of previous studies focus on one type of SOC without another SOC. It's easy for us to understand their physical effect respectively. Out of our interest, we would like to study the competition with SOAM and SLM coupling this letter, we investigate the ground state of two-component Bose gas under SOAM and SLM coupling. In order to find out the physical effect, we firstly to realize the SOAM coupling by a toy model. By real-space boson dynamical-mean-field theory, the ground-state phase diagram of the toy model is obtained .We get vortex phase with different winding number and core-antiferro in Mott. There is normal Superfluid and vortex Superfluid cross Mott region. Utilizing the effective spin model, we obtain Dzyaloshinskii-Moriya(DM) interaction which is from the SOAM coupling. Then, we study the competition of SOAM and SLM coupling with vanishing detuning. It's nice to see the new phases with spiral phase, canted antiferro and stripe phase .

The paper is organized as follows: in section \uppercase\expandafter{\romannumeral2}, we introduce our setup with SOAM coupling, and the extended Bose-Hubbard model. In Section \uppercase\expandafter{\romannumeral3}, we give a detailed description of our RBDMFT approach. Section \uppercase\expandafter{\romannumeral4} covers our results for our model. We summarize with a discussion in Section \uppercase\expandafter{\romannumeral5}.

\section{Model and Hamiltonian}
\begin{figure}
\includegraphics[trim = 0mm 0mm 0mm 0mm, clip=true, width=0.475\textwidth]{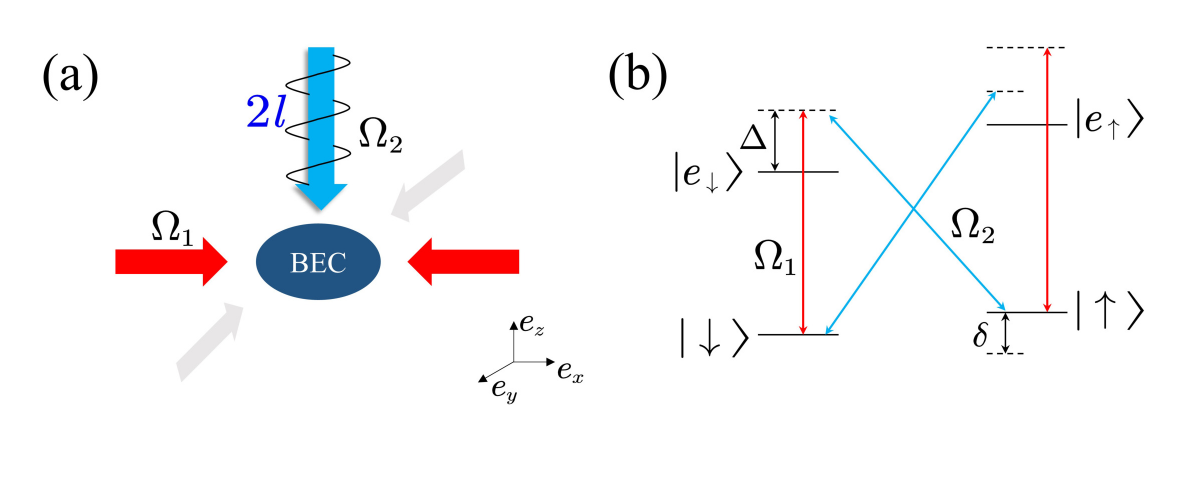}
\caption{(Color online) (a) Sketch of Raman couplings, induced by a plan-wave laser with orbital angular momentum $2l$ in the $z$ direction, and a standing wave in the $x$ direction. To achieve a two-dimensional square lattice, both a standing wave in the $y$-direction and a strong confinement freezing the motional degree of freedom of the atoms in the $z$ direction are added. (b) Atomic level diagram coupled by the pairs of the laser beams $\Omega_{1,2}$.}
\label{1}
\end{figure}
%$\Omega_1$ is Raman beam with optical lattice of $x$ direction and grey beam just make optical lattice in $y$ direction.
We consider two-component bosonic gases trapped in a conventional two-dimensional (2D) square lattice. A plane-wave laser with orbital angular momentum $2l$ is added in the $z$-direction, as shown in Fig.~\ref{1}(a). The two spin states are denoted as $\sigma = \uparrow$ and $\downarrow$, which are coupled by Raman transitions induced by the standing wave with Rabi frequency $\Omega_1(r)$ in the $x$ direction, and the plane-wave laser with Rabi frequency $\Omega_2(r)$ in the $z$ direction~\cite{PhysRevB.96.144517,PhysRevLett.112.086401,PhysRevLett.102.046402,PhysRevLett.110.076401}. In the large detuning limit $\Omega_{1,2}\ll \left|\Delta \right|$, this system can be described by an effective single-particle Hamiltonian (see Appendix A)~\cite{PhysRevLett.110.076401,PhysRevA.91.033630}
\begin{eqnarray}
\label{eq:Ham}
\mathcal{H}_s&=&\frac{p^2}{2m}-\frac{l\hbar}{mr^2}L_z\sigma _z+\frac{l^2\hbar ^2}{2mr^2}+V_{\rm{ext}}(r)\nonumber \\
&+&\frac{\delta}{2}\sigma_z+\Omega^\prime(r)\cos kx\cdot \sigma _x,
\end{eqnarray}
where $L_z$ denotes orbital-angular-momentum operator of atoms along the $z$ direction, $\delta$ the effective Zeeman field, and $\Omega^\prime(r)\cos kx$ the periodic Raman field with $\Omega^\prime(r)=\frac{\Omega_1(r)\Omega_2(r)}{\Delta}$ being the effective Raman Rabi coupling. $V_{\rm{ext}}$ denotes the external trap potential in the $x-y$ plane, and in the following we choose an isotropic hard-wall box potential, which has already been realized experimentally~\cite{2017A,2021Quantum}. %As illustrated in Fig.~\ref{1},we realize SLM coupling along $x$ direction and SOAM coupling simultaneously.

For a sufficiently deep blue-detuned ($\Delta>0$) optical lattice, the single-particle states at each site can be approximated by the lowest-band Wannier function $\omega\left( \boldsymbol{x}-\boldsymbol{R}_j \right)$. In this approximation, the single-particle Hamlitonian~(\ref{eq:Ham}) can be cast into a tight-binding model
\begin{eqnarray}
\label{eq:Ham1}
\mathcal{H}_0&=& -\sum_{\langle i,j\rangle,\sigma}\left({t_\sigma c_{i\sigma}^{\dagger}c_{j\sigma}+it_{ij} (c_{i\uparrow}^{\dagger}c_{j\uparrow}-c_{i\downarrow}^{\dagger}c_{j\downarrow})+ {\rm H.c.}}\right)
\nonumber \\
&+&\underset{i_x}\sum{ (-1)^{i_x}\Omega \left( c_{i_x\uparrow}^{\dagger}c_{i_{x+1}\downarrow}-c_{i_x\uparrow}^{\dagger}c_{i_{x-1}\downarrow} + {\rm H.c.} \right)}
\nonumber \\
&+&\sum_i{V^i_{\rm trap} n_{i\sigma}+m_z\left( n_{i\uparrow}-n_{i\downarrow} \right)},
\end{eqnarray}
where $\langle i,j\rangle$ denotes nearest neighbors between sites $i$ and $j$, and $i_x$ is the site in the $x$ direction. $c_{i\sigma}^{\dagger}$ and $c_{i\sigma}$ are creation and annihilation operators for site $i$ and spin $\sigma$, respectively. $t_\sigma$ denotes conventional hopping amplitudes between nearest neighbors, $m_z$ the Zeeman field, $n_{i\sigma}=c^\dagger_{i\sigma}c_{i\sigma}$ the local density, and $V^i_{\rm trap}$ the external trap with the contribution from centrifugal potential $\frac{l^2\hbar^2}{2mr^2}$ being absorbed. $t_{ij}$ is the nearest-neighbor hopping induced by SOAM coupling, favoring hopping along the azimuthal direction (see Appendix B)~\cite{PhysRevA.69.043609,PhysRevA.74.063606,PhysRevLett.96.060405,PhysRevA.76.043601}
\begin{eqnarray}
\label{hop1}
t_{ij}&=&\int{d{\boldsymbol x}}\omega^{\ast}\left( \boldsymbol{x}-\boldsymbol{R}_i \right) \left( \frac{l\hbar}{mr ^2}L_z \right) \omega\left( \boldsymbol{x}-\boldsymbol{R}_j \right)\nonumber \\
&\approx&\left(\frac{x_iy_j-x_jy_i}{r^{\prime 2}} \right) t_{soc},
\end{eqnarray}
where $t_{soc}=-\frac{l\hbar^2}{dm}\int dx \omega^{\ast}(x-d)\partial_x \omega(x)$ with $d$ being lattice constant. Here, $(x_i, y_i)$ are the coordinates of the $i$th site with the origin at the trap center, and $r^{\prime}$ denotes the lattice spacing between the midpoint of sites $i$, $j$, and the trap center. The Raman-assisted nearest-neighbor spin-flip hopping along the $x$ direction
\begin{eqnarray}
\label{hop2}
\Omega=\int{d{\boldsymbol x}}\Omega^{\prime}(r)\omega^{\ast}( \boldsymbol{x}-\boldsymbol{R}_i ) \left| \cos kx \right|\omega( \boldsymbol{x}-\boldsymbol{R}_j ),
\end{eqnarray}
where the Raman-assisted onsite spin-flip hopping is zero, since atoms are symmetrically localized at the nodes for the blue-detuned lattice potential~\cite{PhysRevLett.110.076401,PhysRevB.96.144517}.

For a deep lattice, interaction effects should be included. The $s$-wave contact interaction is given by
\begin{eqnarray}
\label{energy}
\mathcal{H}_{\rm int}&=&\underset{i,\sigma\sigma^{\prime}}\sum \frac{1}{2}U_{\sigma \sigma^{\prime}}n_{i\sigma}\left(n_{i\sigma^{\prime}}-\delta_{\sigma\sigma^{\prime}}\right),
\end{eqnarray}
where $U_{{\uparrow \uparrow},{\downarrow \downarrow}}$ and $U_{\uparrow \downarrow}$ denote the intra- and interspecies interactions, respectively. Additionally, we limit present study to the situations in which the interactions are repulsive and two hyperfine components are miscible with $U_{\uparrow \uparrow}=U_{\downarrow \downarrow}\equiv1.01U_{\uparrow \downarrow}$ and $t\equiv t_\uparrow=t_\downarrow$, which is a good approximation for two-hyperfine-state mixtures of a $^{87}$Rb gas~\cite{PhysRevLett.103.245301}. Thus, the total Hamiltonian of our system reads
\begin{eqnarray}
\label{eq:Hubbard}
\mathcal{H}=\mathcal{H}_0+\mathcal{H}_{\rm int}-\underset{i\sigma}\sum \mu_{\sigma} n_{i\sigma},
\end{eqnarray}
where $\mu_{\sigma}$ is the chemical potential for component $\sigma$. Due to the competition between SOAM coupling and Raman-induced hopping, it is expected that various many-body phases develop in the strongly interacting many-body system described by Eq.~(\ref{eq:Hubbard}). To resolve these quantum phases, we apply real-space bosonic dynamical mean-field theory (RBDMFT), to obtain the complete phase diagrams. In the following, we set $U_{\uparrow \downarrow}\equiv1$ and optical lattice spacing $d\equiv 1$ as the units of energy and length, respectively. We focus on the lower filling case with filling $n_i=n_{i\uparrow}+n_{i\downarrow}=1$ in the Mott regime (the total particle number $N=\sum_i n_i=330$), and the lattice size $N_{\rm lat}=24\times 24$.

{\section{Method}
To resolve the long-range order, we utilize bosonic dynamical mean-field theory (BDMFT) to calculate many-body ground states of the system described by Eq.~(\ref{eq:Hubbard}). By neglecting non-local contributions to the self-energy within BDMFT~\cite{RevModPhys.68.13}, the $N$-site lattice problem can be mapped to $N$ single-impurity models interacting with two baths, which correspond to condensing and normal bosons, respectively~\cite{PhysRevLett.121.093401,PhysRevB.84.144411,PhysRevA.87.051604,PhysRevB.77.235106}. By a self-consistency condition, we can finally obtain the physical information of the $N$-site model. Note here that, in a real-space system without lattice-translational symmetry, the self-energy is lattice-site dependent, i.e. $\Sigma_{i,j}=\Sigma_i \delta_{ij}$ with $\delta_{ij}$ being a Kronecker delta, which motivates us to utilize a real-space version of BDMFT~\cite{Snoek_2008,chatterjee2019real,PhysRevLett.100.056403}.

In RBDMFT, our challenge is to solve the single-impurity model, and the physics of site $i$ is given by the local effective action $\mathcal{S}^{(i)}_{imp}$. Following the standard derivation~\cite{RevModPhys.68.13}, we can write down the effective action for impurity site $i$, which is described by
\begin{widetext}
\begin{eqnarray}
\label{eq:action}
\mathcal{S}_{imp}^{(i)}&=&-\int_0^{\beta}{d}\tau d\tau^{\prime}\underset{\sigma \sigma ^{\prime}}\sum \boldsymbol{c}_{\sigma}^{\left( i \right)}\left( \tau \right) ^{\dagger}\boldsymbol{\mathcal{G}} _{\sigma \sigma ^{\prime}}^{\left( i \right)}\left( \tau -\tau ^{\prime} \right) ^{-1}\boldsymbol{c}_{\sigma ^{\prime}}^{\left( i \right)}\left( \tau ^{\prime} \right) +\int_0^{\beta}{d}\tau \frac{1}{2}\underset{\sigma \sigma ^{\prime}}\sum U_{\sigma \sigma ^{\prime}}{n^{\left( i \right)}}_{\sigma}\left( \tau \right) \left( {n^{\left( i \right)}}_{\sigma ^{\prime}}\left( \tau \right) -\delta _{\sigma \sigma^{\prime}} \right) \nonumber \\
&+& \frac{1}{z}\int_0^{\beta}{d}\tau \left( -\underset{\langle i,j\rangle,\sigma}\sum t_{\sigma}\left( \boldsymbol{c}_{\sigma}^{\left( i \right)}\left( \tau \right) \left( \begin{array}{c}	\phi _{j,\sigma}^{\left( i \right)}\left( \tau \right) ^{\ast}\\	\phi _{j,\sigma}^{\left( i \right)}\left( \tau \right)\\\end{array} \right) \right) +it_{ij}\left( \boldsymbol{c}_{\uparrow}^{\left( i \right)}\left( \tau \right) \left( \begin{array}{c}	\phi _{j,\uparrow}^{\left( i \right)}\left( \tau \right) ^{\ast}\\	-\phi _{j,\uparrow}^{\left( i \right)}\left( \tau \right)\\\end{array} \right)  -\boldsymbol{c}_{\downarrow}^{\left( i \right)}\left( \tau \right) \left( \begin{array}{c}	\phi _{j,\downarrow}^{\left( i \right)}\left( \tau \right) ^{\ast}\\	-\phi _{j,\downarrow}^{\left( i \right)}\left( \tau \right)\\\end{array} \right) \right) \right.   \nonumber \\
&+&\left. \underset{i_x,\sigma \ne \sigma ^{\prime}}{\sum}(-1)^{i_x}\Omega\left( c_{\sigma}^{\left( i_x \right)}\left( \tau \right) ^{\ast}\left( {\phi ^{\left( i_x \right)}}_{i_x+1,\sigma ^{\prime}}\left( \tau \right) -{\phi ^{\left( i_x \right)}}_{i_x-1,\sigma '}\left( \tau \right) \right) +{c^{\left( i_x \right)}}_{\sigma}\left( \tau \right) \left( \phi _{i_x+1,\sigma ^{\prime}}^{\left( i_x \right)}\left( \tau \right) ^{\ast}-\phi _{i-1,\sigma '}^{\left( i_x \right)}\left( \tau \right) ^{\ast} \right) \right) \right. \nonumber \\
&+&\left.\underset{i,\sigma \ne \sigma ^{\prime}}{\sum}V^{\left( i \right)}_{\rm trap}{n^{\left( i \right)}}_{\sigma}\left( \tau \right) +m_z\left( {n^{\left( i \right)}}_{i\sigma}\left( \tau \right) -{n^{\left( i \right)}}_{i\sigma '}\left( \tau \right) \right) \right).
\end{eqnarray}
\end{widetext}
Here, $\boldsymbol{\mathcal{G}}^{(i)}_{\sigma \sigma^{\prime}}\left( \tau-\tau^{\prime} \right)$ is a local non-interacting propagator interpreted as a dynamical Weiss mean field which simulates the effects of all other sites. To shorten the formula, the Nambu notation is used $\boldsymbol{c}^{(i)}_{\sigma}(\tau)\equiv(c^{(i)}_{\sigma}(\tau),c^{(i)}_{\sigma}(\tau)^{\ast})$. The parameter $z$ is the lattice coordination, which is treated as a control parameter within RBDMFT. The terms up to subleading order are included in the effective action. The static bosonic mean-fields are defined in terms of the bosonic operator $c_{j\sigma}$ as
\begin{eqnarray}
\phi^{(i)}_{j\sigma}\left(\tau \right)=\left< c^{(i)}_{j\sigma} \left( \tau \right)\right>_0,
\end{eqnarray}
where $\left< ... \right>_0$ means the expectation value in the cavity system without the impurity site.

Instead of solving the effective action directly, we normally turn to the Hamiltonian representation, i.e. Anderson impurity Hamiltonian~\cite{PhysRevB.80.245109,PhysRevB.84.144411}. By exactly diagonalizing the Anderson impurity Hamiltonian with a finite number of bath orbitals~\cite{RevModPhys.68.13,bath}, we can finally obtain the local propagator
\begin{eqnarray}
{G}^{(i)}_{\sigma\sigma^{\prime},imp}\left(\tau,\tau^{\prime} \right )=-\left< T \boldsymbol{c}^{(i)}_{\sigma}(\tau) \boldsymbol{c}^{(i)}_{i\sigma^{\prime}}(\tau^{\prime})^{\dagger}\right>_{\mathcal{S}^{(i)}_{imp}}.
\end{eqnarray}
Next, we utilize the Dyson equation to obtain site-dependent self-energies in the Matsubara frequency representation
\begin{eqnarray}
\label{Dyson}
{\Sigma}^{(i)}_{\sigma\sigma^{\prime},imp}(i\omega_n)={\mathcal{G}}^{(i)}_{\sigma\sigma^{\prime}}(i\omega_n)^{-1}-{G}^{(i)}_{\sigma\sigma^{\prime}}(i\omega_n).
\end{eqnarray}
In the framework of RBDMFT, we assume that the impurity self-energy $\Sigma_{imp}(i\omega_n)$ is local (momentum-independent) and coincides with lattice self-energy $\Sigma_{lattice}(i\omega_n)$, whose assumption is exact in infinite dimensions and good approximations in higher dimensions~\cite{RevModPhys.68.13}. Finally, we employ the Dyson equation in the real-space representation to obtain the interacting lattice Green's function
\begin{eqnarray}
\label{lattice_G}
\boldsymbol{G}_{\sigma\sigma^{\prime},lattice}=\frac{1}{i\omega_n+\boldsymbol{\mu}-\boldsymbol{\varepsilon}-\boldsymbol{\Sigma}_{imp}(i\omega_n)},
\end{eqnarray}
where boldface quantities denote matrices with site-dependent elements. $\boldsymbol{\varepsilon}$ denotes a matrix with the elements being nearest-neighbor hopping amplitudes for a given lattice structure, $\boldsymbol{\mu}$ represents the onsite hopping amplitudes with the external trap, and $\boldsymbol{\Sigma}_{imp}(i\omega_n)$ denotes the self-energy. The self-consistency RBDMFT loop is closed by the Dyson equation to obtain a new local non-interacting propagator ${\mathcal{G}^{i}_{\sigma\sigma^\prime}}$. These processes are repeated until the desired accuracy for superfluid order parameters and noninteracting Green's functions is obtained.

%The self-consistency loop is solved as follow: starting from solving the Anderson Hamiltonian the same effective action with our system by exact diagonalization to obtain the impurity Green's functions. The self-energy is obtained by Dyson equation Eq.~(\ref{Dyson}).Next, we get interacting lattice Green's function according to Eq.~(\ref{lattice_G}).Finally, the Weiss mean-field is obtained from Dyson equation. The self-consistency equations are closed. This procedure is repeated until convergence is reached.

{\section{Results}
\begin{figure}
\includegraphics[trim = 0mm 0mm 0mm 0mm, clip=true, width=0.475\textwidth]{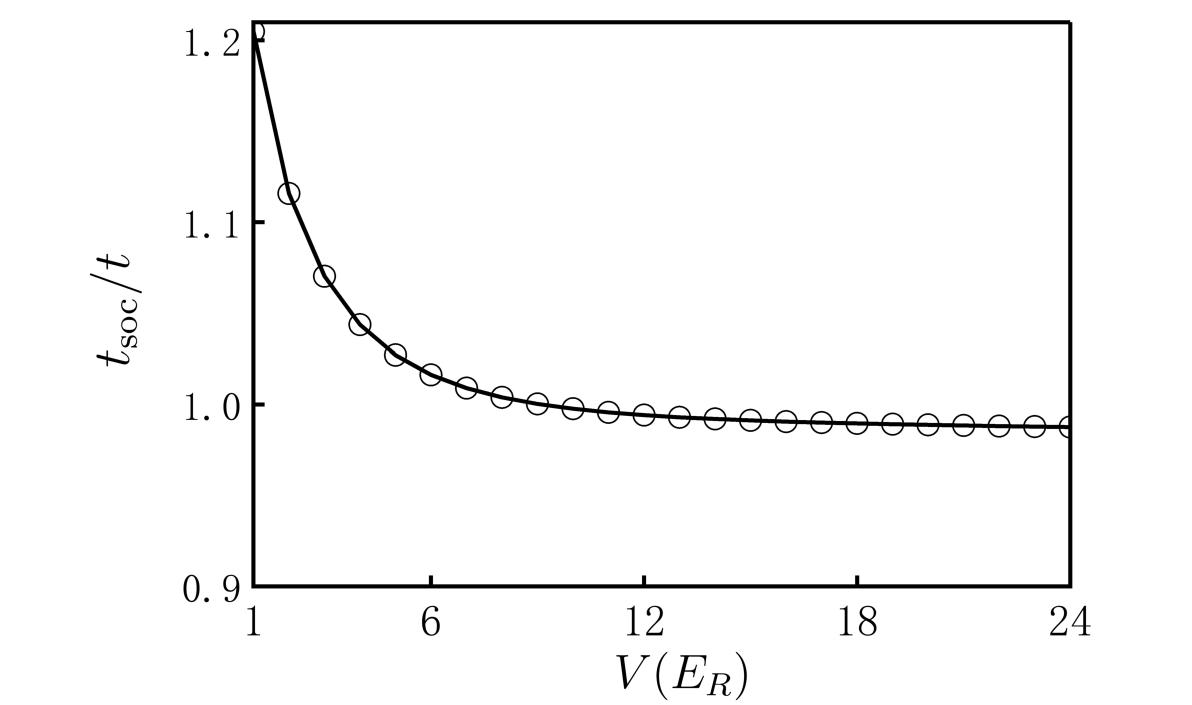}
\caption{(Color online) Nearest-neighbor hopping amplitudes ${t_{soc}}/{t}$ as a function of lattice depth $V$. In the regime with $V>5 \,E_R$, ${t_{soc}}/{t}$ scales roughly linearly with the lattice depth $V$, where $E_R$ is the recoil energy. We choose the orbital angular momentum $l=1$.}
\label{fig_3}
\end{figure}

{\subsection{Spin-orbital-angular-momentum coupling}
Before exploring the whole model, described by Eq.~(\ref{eq:Hubbard}), we first discuss the competition between conventional nearest-neighbor hopping $t$ and orbital-angular-momentum-induced hopping $t_{soc}$. As shown in Fig.~\ref{fig_3}, the orbital-angular-momentum-induced hopping can be the order of the conventional one even for $l=1$, where the hopping amplitudes are obtained from band-structure simulations~\cite{band_structure}. %First at all, we want to find out the effect of SOAM coupling term. According the Eq.~(\ref{hop1}),when we fix the depth of optical lattice, $t_{soc}$ is quantized value for the OAM $l$ and has a relationship with $t$, as Fig.~\ref{fig_3}.In this part, we consider $t_l$ a continuous value for convenience. By We have a toy model now, reads
By neglecting the Raman-induced spin-flip hopping $\Omega$, Eq.~(\ref{eq:Hubbard}) is reduced to
\begin{eqnarray}
\label{eq:SOAM}
H&=&-\sum_{\langle i,j\rangle,\sigma}\left(tc_{i\sigma}^{\dagger}c_{j\sigma} + it_{ij} (c_{i,\uparrow}^{\dagger}c_{j,\uparrow}-c_{i,\downarrow}^{\dagger}c_{j,\downarrow})+ {\rm H.c.}\right) \nonumber \\
&+&\underset{i,\sigma\sigma^{\prime}}\sum \frac{1}{2}U_{\sigma \sigma^{\prime}}n_{i\sigma}\left(n_{i\sigma^{\prime}}-\delta_{\sigma\sigma^{\prime}}\right)+V^{i}_{\rm trap}n_{i,\sigma} -\mu_{i\sigma} n_{i\sigma},\nonumber\\
\end{eqnarray}
with $m_z=0$.
\begin{figure}
\includegraphics[trim = 0mm 0mm 0mm 0mm, clip=true, width=0.5\textwidth]{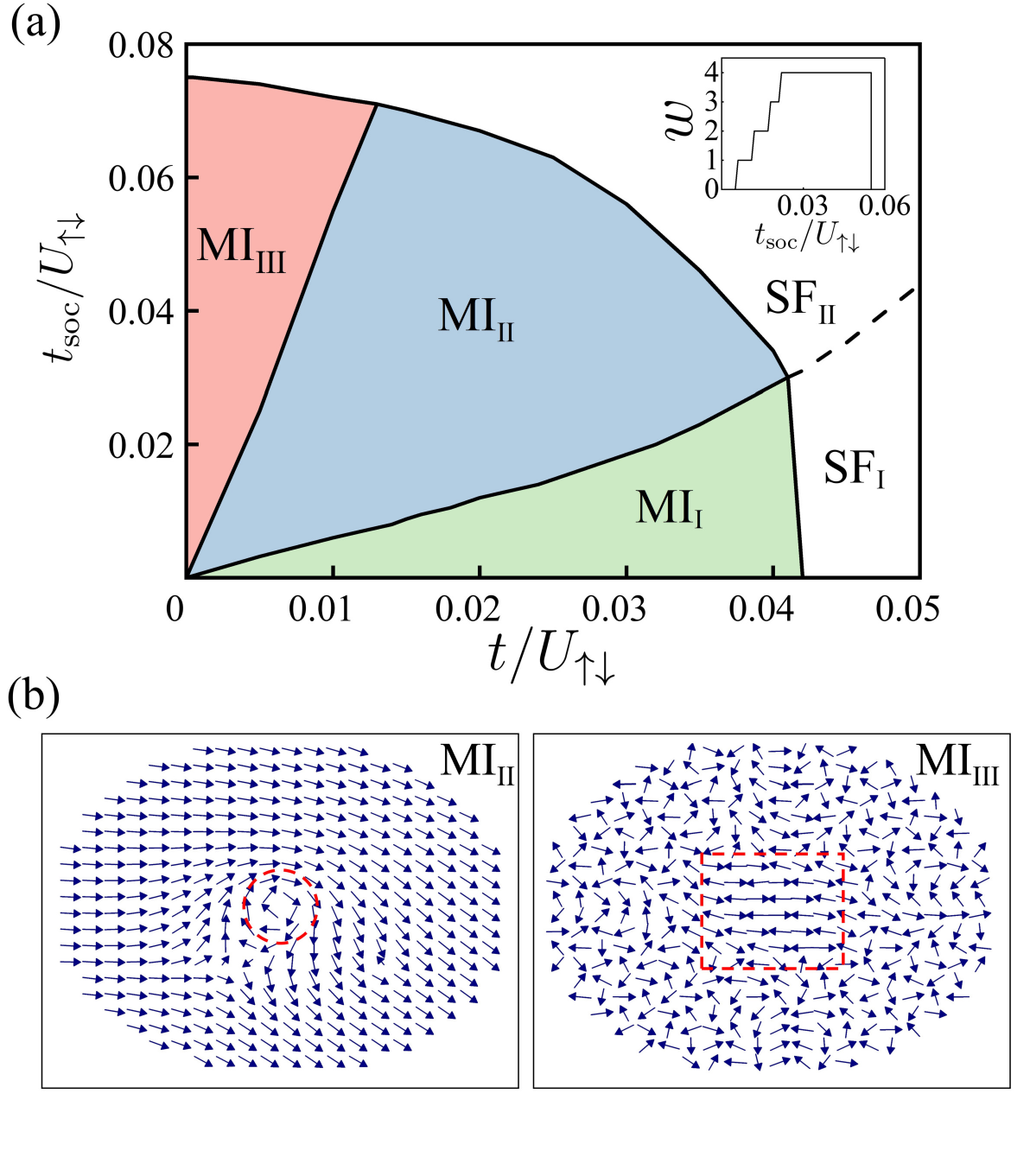}
\caption{(Color online) (a) Many-body phase diagram of two-component ultracold bosonic gases in a square lattice in the presence of SOAM coupling, described by Eq.~(\ref{eq:SOAM}). The system favors three Mott phases with ferromagnetism ($\rm MI_I$), spin-vortex ($\rm MI_{II}$), and composite spin-vortex with antiferromagnetic core ($\rm MI_{III}$). In the superfluid regime, two quantum phases appear, denoted as conventional superfluid $(\rm SF_{\uppercase \expandafter{\romannumeral1}})$ and rotating superfluid $(\rm SF_{\uppercase \expandafter{\romannumeral2}})$. Inset: winding number $w$ as a function of $t_{soc}$ with $t=0.015$. (b) Contour plots of spin textures for phases $\rm MI_{II}$ and $\rm MI_{III}$ in the Mott-insulating regime. The interactions $U_{\uparrow \uparrow}=U_{\downarrow \downarrow}=1.01U_{\uparrow \downarrow}$. }
\label{fig_2}
\end{figure}

\begin{figure*}
\includegraphics[trim = 0mm 0mm 0mm 0mm, clip=true, width=1\textwidth]{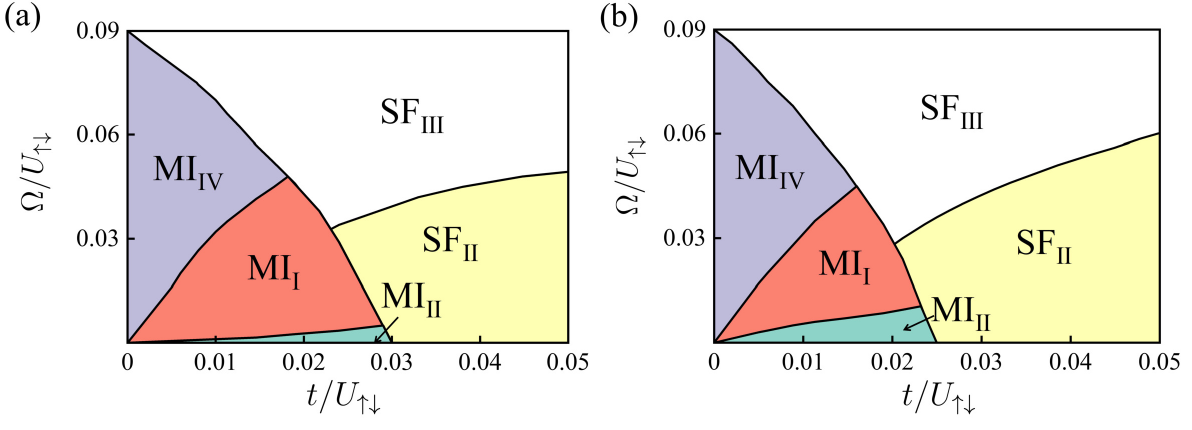}
\caption{(Color online) Many-body phase diagrams of two-hyperfine-state mixtures of a $^{87}$Rb gas in a square lattice, described by Eq.~(\ref{eq:Hubbard}), in the presence of SOAM coupling and Raman-induced spin-flip hopping for orbital angular momenta $l=1$ (a) and $l=2$ (b). The system favors Mott phases with ferromagnetic ($\rm MI_{ \uppercase\expandafter{\romannumeral1}}$), vortex ($\rm MI_{\uppercase\expandafter{\romannumeral2}}$), and canted-antiferromagnetic ($\rm MI_{ \uppercase\expandafter{\romannumeral4}}$) orders, and superfluid phases with vortex ($\rm SF_{ \uppercase\expandafter{\romannumeral2}}$) and stripe ($\rm SF_{ \uppercase\expandafter{\romannumeral3}}$) textures. The interactions $U_{\uparrow\uparrow}/U_{\uparrow\downarrow}=U_{\downarrow\downarrow}/U_{\uparrow\downarrow}=1.01$, and the effective Zeeman field $m_z=0$. }
\label{fig_4}
\end{figure*}

As shown in Fig.~\ref{fig_2}, a many-body phase diagram is shown as a function of hopping amplitudes $t$ and $t_{soc}$ for interactions $U_{\uparrow \uparrow}=U_{\downarrow \downarrow}=1.01U_{\uparrow \downarrow}$, based on RBDMFT. To distinguish quantum phases, we introduce superfluid order parameters $\phi_\sigma$, pseudospin operators $S^z_i=\frac{1}{2}({b}^{\dagger}_{i,\uparrow}b_{i,\uparrow}-{b}^{\dagger}_{i,\downarrow}b_{i,\downarrow})$, $S^x_i=\frac{1}{2}({b}^{\dagger}_{i,\uparrow}b_{i,\downarrow}+{b}^{\dagger}_{i,\downarrow}b_{i,\uparrow})$ and $S^y_i=\frac{1}{2i}(b^{\dagger}_{i,\uparrow}b_{i,\downarrow}-{b}^{\dagger}_{i,\downarrow}b_{i,\uparrow})$, and winding number $w=\frac{1}{2\pi}\sum_{\mathcal{C}} {\rm arg} (M^\ast_i M_j)$ with $M_i=S^x_i+iS^y_i$ and $\mathcal{C}$ being a closed loop around the center of the trap~\cite{PhysRevA.101.013427,zhang2017direct}. We observe five different quantum phases. When $t_{soc} \ll t$, the system demonstrates a ferromagnetic phase ($\rm MI_I$), identical with the system without SOAM coupling. With the growth of $t_{soc}$, a spin-vortex phase appears in the Mott-insulating regime with $w\neq0$ ($\rm MI_{II}$), since the growth of $t_{soc}$ is equivalent to the growth of orbital angular momentum $l$. As shown in Fig.~\ref{fig_2}(b), SOAM-induced spin rotation appears around the trap center with winding number $w=1$, indicating that the spin rotates slowly with the corresponding response being mainly around the center of the trap. The physical reason is that the SOAM-induced hopping is site-dependent, and pronounced around the trap center, as indicated by Eq.~(\ref{hop1}). Further increasing $t_{soc}$, the winding number $w$ grows as well, as shown in the inset of Fig.~\ref{fig_2}(a), and finally we observe the whole system rotating in the regime $t_{soc}\gg t$ ($\rm MI_{III}$). Interestingly, this spin-vortex phase is actually a composite vortex defect, which supports a {\it nonrotating} core of antiferromagnetic spin texture, with the nearest-neighbor spins being antiparallel in the trap center, as shown in Fig.~\ref{fig_2}(b).
%In other words, the winding of the spins is suppressed in the core region, for a bigger orbital angular momentum. %the system's core would rotate fatly with the result of antiferromagnets near the center of rotation.

To understand the underlying physics in the Mott regime, we treat the hopping as perturbations and derive an effective exchange model at half filling. With defining the projection operators $\mathcal{P}$ and $\mathcal{Q} = 1 - \mathcal{P}$, we can project the system into the Hilbert space consisting of both singly occupied sites and the states being at least one site with double occupation, and obtain an effective exchange model $H_{eff}=-\mathcal{P}H_t\mathcal{Q}(\frac{1}{\mathcal{Q}H_U\mathcal{Q}-E})\mathcal{Q}H_t\mathcal{P}$~\cite{PhysRevLett.111.205302,pinheiro2016multi,mila2011strong}. The effective exchange model is finally given by:
\begin{eqnarray}
\label{eff_soc_only}
H_{eff}&=&\sum_{\langle i,j\rangle}J_z S_{i}^{z}S_{j}^{z}+J\left( S_{i}^{x}S_{j}^{x}+S_{i}^{y}S_{j}^{y} \right) \nonumber \\
&+&D\left( S_i\times S_j \right) _z.
\end{eqnarray}
Here, $J_z=-4(\frac{2}{U}-\frac{1}{U_{\uparrow \downarrow}})(t^2+t^2_{ij})$, $J=-\frac{4(t^2-t^2_{ij})}{U_{\uparrow \downarrow}}$, and $D=-\frac{8tt_{ij}}{U_{\uparrow \downarrow}}$. The details of derivation are given in the Appendix C.

In the absence of SOAM coupling, this effective model is reduced to the conventional XXZ model, where the system prefers ferromagnetic and anti-ferromagnetic orders~\cite{PhysRevLett.91.090402,PhysRevLett.90.100401,2003Phase,PhysRevB.72.184507}. In the presence of SOAM coupling, a Dzyaloshinskii-Moriya term~\cite{fert2013skyrmions,luo2014effect} appears in the $z$ direction. This term competes with the normal Heisenberg exchange interactions, resulting spin-vortex defects in the Mott-insulating regime. Interestingly, the Heisenberg exchange term $J$ also depends on the SOAM-induced hopping $t_{soc}$, and dominates in the regime $t_{soc}\gg t$, resulting an antiferromagnetic texture. This texture is consistent with our numerical results, as shown in Fig.~\ref{fig_2}(b). We remark that the SOAM-induced Dzyaloshinskii-Moriya term $D$ preserves rotational symmetry with spin texture rotating along the azimuthal direction, in contrast to the spin-linear-momentum coupling by breaking lattice-translational symmetry~\cite{PhysRevLett.109.085302,PhysRevA.92.023630,PhysRevA.78.023616,PhysRevA.84.053632,PhysRevB.86.155101}.
%we see third term is DM interaction of $z$ direction from SOAM. In magnetic system, DM interaction is an essential term in models of spiral order and multiferroic effects in general, which is usually induced by a lack of inversion symmetry and strong spin-orbit coupling.~\cite{fert2013skyrmions,luo2014effect}.It's that this term bring the vortex state into our system.

%Actually, as Fig.~\ref{fig_3} showing, $t_{soc}$ and $t$ is related to the depth of optical lattice. We can adjust the proportion of $t_{soc}/t$ by OAM $l$. According to the results of RBDMFT, we can get easily the vortex phase in Mott and Superfluid in $l=1$.And $l\ge 6$ we are into core-antiferro phase. We would compute by setting the depth of optical lattice to control $t$ and $t_{soc}$ next.

With the increase of hopping amplitudes, atoms delocalize and the superfluid phase appears. We characterize the superfluid phase with superfluid order parameters $\phi_\sigma$. In the superfluid region, we observe two quantum many-body phases, with one being a phase with phase rotating ($\rm SF_{\uppercase \expandafter{\romannumeral2}}$), and the other with conventional phase ($\rm SF_{\uppercase \expandafter{\romannumeral1}}$).

\begin{figure}
\includegraphics[trim = 0mm 0mm 0mm 0mm, clip=true, width=0.475\textwidth]{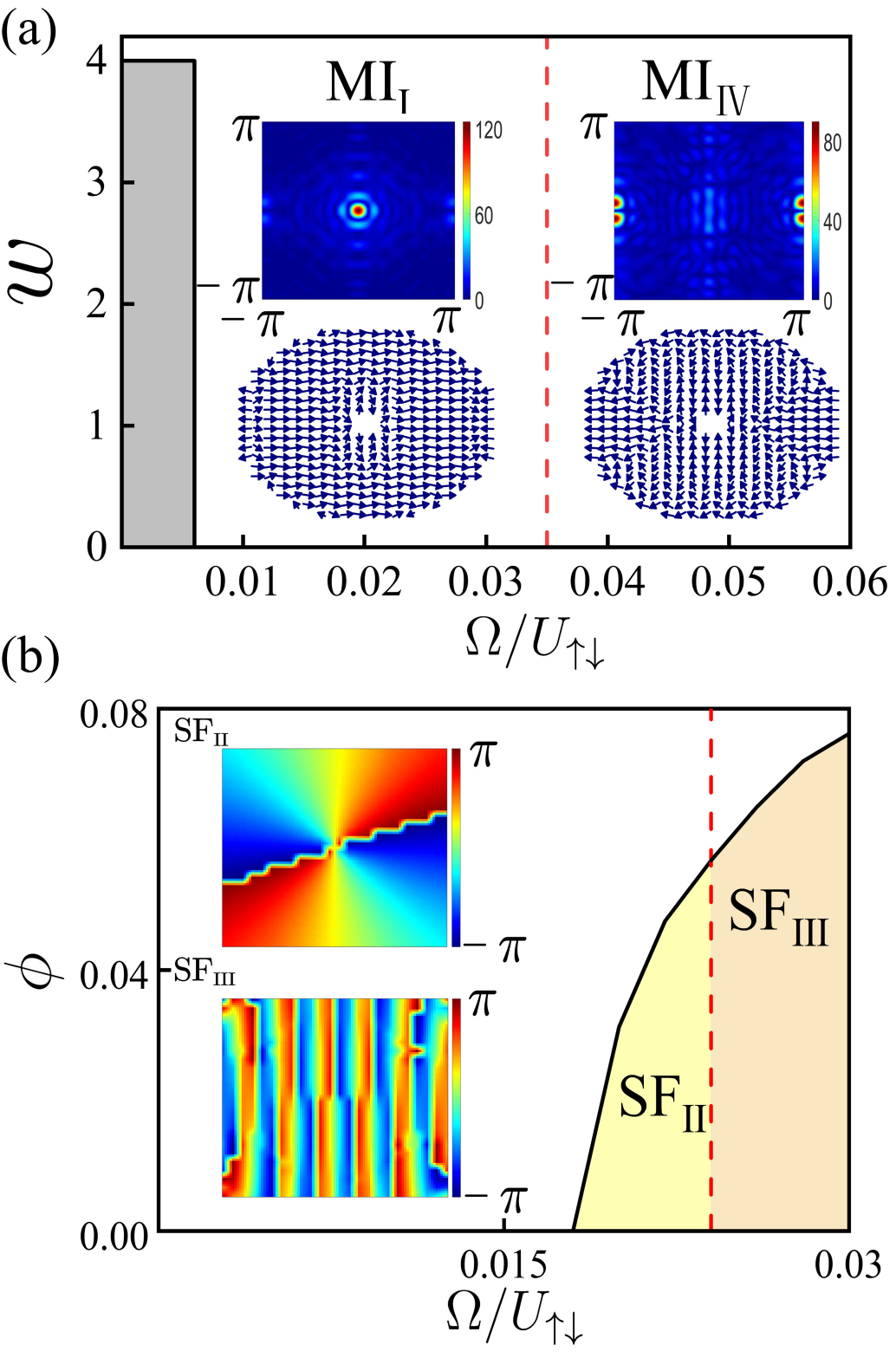}
\caption{(Color online) Phase transitions as a function of Raman-induced spin-flip hopping for different lattice depths $V=14\, E_R$ ($t \approx0.011$) (a) and $V=11.5\, E_R$ ($t \approx0.022$) (b). Inset: spin structure factor (upper) and real-space spin texture (lower) for different phases (a), and local phases of superfluid order parameter for the spin-$\uparrow$ component (b). The interactions $U_{\uparrow\uparrow}/U_{\uparrow\downarrow}=U_{\downarrow\downarrow}/U_{\uparrow\downarrow}=1.01$, and the orbital angular momentum $l=2$.  }
%The left of red dash line is spin and spin structure of spiral order with $\omega=0$. The grey area is vortex phase with $\omega=4$.(b) The phase transition in Superfluid. The solid line is order parameter of superfluid $\Phi$. Inset: The top plot is phase of $\Phi$ in vortex($\rm SF_{ \uppercase\expandafter{\romannumeral2}}$).The top plot is phase of $\Phi$ in vortex($\rm SF_{ \uppercase\expandafter{\romannumeral3}}$). The interaction is the same as Fig.~\ref{fig_2}
\label{fig_5}
\end{figure}
{\subsection {Interplay of spin-orbital-angular-momentum coupling and Raman-induced spin-flip hopping}}
Now we turn to study the whole system, described by Eq.~(\ref{eq:Hubbard}), and focus on the stability of spin-vortex texture in the strongly interacting regime. Generally, SOAM coupling preserves rotational symmetry and favors spin-vortex defects~\cite{PhysRevA.102.013316,PhysRevResearch.2.033152}, whereas the one-dimensional Raman-induced spin-flip hopping prefers the stripe phase and breaks translational symmetry~\cite{PhysRevLett.105.160403,Zhai_2015}. It is expected that more exotic many-body phases appear, due to the competition between SOAM coupling and Raman-induced spin-flip hopping. Here, we choose two hyperfine states of a $^{87}$Rb gas as examples, where all the Hubbard parameters are obtained from band-structure simulations~\cite{band_structure}. To emphasize the influence of SOAM coupling, we consider the orbital angular momenta $l=1$ and $l=2$. Note here that $t\approx t_{soc}$ for orbital angular momentum $l=1$ in the deep lattice, as shown in Fig.~\ref{fig_3}. %To study the spin miscible regime, we choose the scattering length with $U_{\uparrow\uparrow}/U_{\uparrow\downarrow}=U_{\downarrow\downarrow}/U_{\uparrow\downarrow}=1.01$~\cite{PhysRevLett.103.245301}.

Rich phases are found in Fig.~\ref{fig_4}, including Mott-insulating phases with ferromagnetic ($\rm MI_{ \uppercase\expandafter{\romannumeral1}}$), vortex ($\rm MI_{\uppercase\expandafter{\romannumeral2}}$), and canted-antiferromagnetic ($\rm MI_{ \uppercase\expandafter{\romannumeral4}}$) orders~\cite{PhysRevMaterials.4.044405}, and superfluid phases with vortex ($\rm SF_{ \uppercase\expandafter{\romannumeral2}}$) and stripe ($\rm SF_{ \uppercase\expandafter{\romannumeral3}}$) patterns. In the limit $\Omega \ll t_{soc}$, the many-body phases develop spin-vortex textures ($\rm MI_{\uppercase\expandafter{\romannumeral2}}$ and $\rm SF_{ \uppercase\expandafter{\romannumeral2}}$), whereas the system prefers density-wave orders ($\rm MI_{ \uppercase\expandafter{\romannumeral4}}$ and $\rm SF_{ \uppercase\expandafter{\romannumeral3}}$) in the limit $\Omega\gg t_{soc}$. This conclusion is consistent with our general discussion above, as a result of the interplay of SOAM coupling and Raman-induced spin-flip hopping. Note here that the region of the spin-vortex phase is enlarged for larger orbital angular momentum, as shown in Fig.~\ref{fig_4}(b), indicating large opportunity for observing this spin texture for larger orbital angular momentum.
%are For $l=1$,the new phases in Mott exists only small region as showing in Fig.~\ref{fig_4}(a).With a larger OAM transfer $l=2$,the region is changed such as Fig.~\ref{fig_4}(b).

To characterize these different phases, we choose winding number, real-space spin texture, spin-structure factor $S_{\vec{q}}=\left|\vec{S}_ie^{i\vec{q}\cdot \vec{r}_i} \right| $ ~\cite{PhysRevLett.109.085302}, and local phase of superfluid order parameter, as shown in Fig.~\ref{fig_5}. Here, we choose the orbital angular momentum $l=2$, and different lattice depths $V=14\, E_R$ with hopping $t\approx0.011$ [Fig.~\ref{fig_5}(a)], and $V=11.5\,E_R$ with $t \approx0.022$ [Fig.~\ref{fig_5}(b)]. For small $\Omega$, a spin-vortex phase ($\rm MI_{ \uppercase\expandafter{\romannumeral2}}$) develops with winding number $w=4$, as shown in Fig.~\ref{fig_5}(a). Increasing $\Omega$, the spin texture changes to ferromagnetic ($\rm MI_{ \uppercase\expandafter{\romannumeral1}}$) and canted-antiferromagnetic ($\rm MI_{ \uppercase\expandafter{\romannumeral4}}$) textures with vanishing winding number, which are characterized both by magnetic spin-structure factor $S_{\vec{q}}$ and real-space spin texture, as shown in the inset of Fig.~\ref{fig_5}(a). We remark here that the $\rm MI_{ \uppercase\expandafter{\romannumeral4}}$ }phase possesses spin-density-wave in $S_z$, ferromagnetic order in $S_x$, and antiferromagnetic order in $S_y$. The local phase of superfluid order parameter is shown in Fig.~\ref{fig_5}(b). For small $\Omega$, a nonzero winding number of the local phase develops in the vortex superfluid ($\rm SF_{II}$). With $\Omega$ larger, we find the local phase demonstrates a stripe order instead ($\rm SF_{III}$).
%It's nothing but when $\Omega$ is small that the system is vortex state in Mott. Under the rise of $\Omega$, we see a different phase that the vortex state become a spiral order along $y$ direction. With $\Omega \gg t$, there is a new phase called canted antiferro phase(CAF)~\cite{PhysRevMaterials.4.044405,morrish1994canted}.It's a new phase with ferromagnetism of $S_x$ and antiferromagnetic of $S_y$.
%In superfluid, we also discover similar phenomena with vortex and stripe phases. With small $\Omega$, it's just like what we get without SLM coupling. However, the influence appears with $\Omega$ larger, we find the phase of $\Phi$ becomes stripe. It's said the system become stripe phase. The system has rotational symmetry in vortex phase. With the SLM coupling, system rotational symmetry is broken. where the left (right) plot is for the spin-$\downarrow$ ($\uparrow$) component for the $\rm SF_{ \uppercase\expandafter{\romannumeral2}}$ phase

\begin{figure}
\includegraphics[trim = 0mm 0mm 0mm 0mm, clip=true, width=0.475\textwidth]{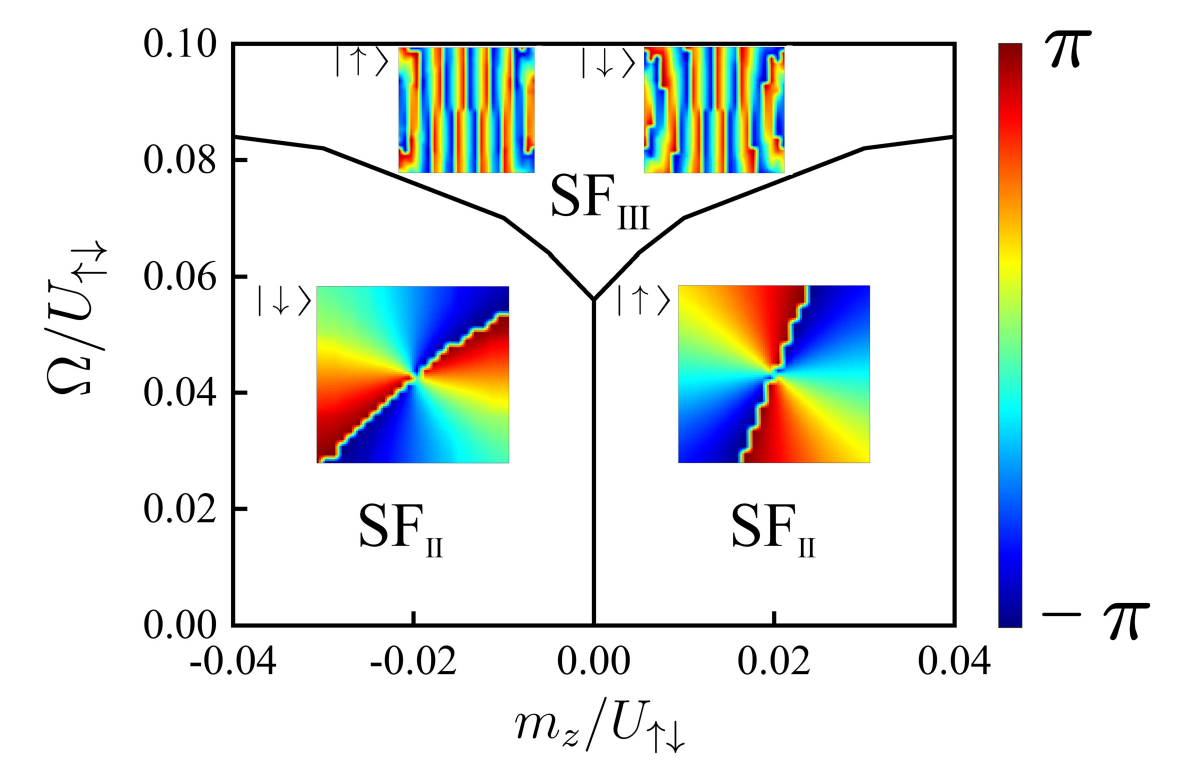}
\caption{(Color online) Many-body phase diagram of two-hyperfine-state mixtures of a $^{87}$Rb gas in a square lattice with the depth $V=9\, E_R$ $(t\approx 0.045)$, as a function of Raman-induced spin-flip hopping $\Omega$ and effective magnetic field $m_z$. Inset: local phases of superfluid order parameters for different phases. The interactions $U_{\uparrow\uparrow}/U_{\uparrow\downarrow}=U_{\downarrow\downarrow}/U_{\uparrow\downarrow}=1.01$, and the orbital angular momentum $l=2$. }
\label{fig_6}
\end{figure}
To understand the physical phenomena in the Mott regime, we derive an effective exchange model of the system at half filling, described by Eq.~(\ref{eq:Hubbard}),
\begin{eqnarray}
\label{eff2}
H_{eff}&=&\underset{\langle i_x,j_x\rangle }\sum J^{\prime}_z S_{i_x}^{z}S_{j_x}^{z}+J^{\prime}_x S_{i_x}^{x}S_{j_x}^{x}+J^{\prime}_y S_{i_x}^{y}S_{j_x}^{y}\nonumber \\
&+&\underset{\langle i_y,j_y\rangle}\sum J_z S_{i_y}^{z}S_{j_y}^{z}+J\left( S_{i_y}^{x}S_{j_y}^{x}+S_{i_y}^{y}S_{j_y}^{y} \right) \nonumber \\
&+&\sum_{\langle i,j\rangle}D\left( S_i\times S_j \right)_z,
\end{eqnarray}
where $J_z^{\prime}=-4\left( \frac{2\left( t^2+t_{ij}^{2} \right)}{U}+\frac{\Omega ^2}{U_{\uparrow \downarrow}}-\frac{\left( t^2+t_{ij}^{2} \right)}{U_{\uparrow \downarrow}}-\frac{2\Omega ^2}{U} \right)$, $J_x^{\prime}=-4\left( \frac{\left( t^2-t_{ij}^{2} \right)}{U_{\uparrow \downarrow}}+\frac{\Omega ^2}{U_{\uparrow \downarrow}} \right) $, and $J_y^{\prime}=-4\left( \frac{\left( t^2-t_{ij}^{2} \right)}{U_{\uparrow \downarrow}}-\frac{\Omega ^2}{U_{\uparrow \downarrow}} \right) $. We observe that the Raman-induced spin-flip hopping does not influence Dzyaloshinskii-Moriya interactions, but induce an anisotropy for Heisenberg exchange interactions in the $x$ and $y$ directions. When $\Omega$ is large enough, the Raman-induced hopping can induce $J^\prime_{y,z}$ to be positive and $J^\prime_x$ negative. It indicates that a spin-density-wave and canted-antiferromagnetic order develops for large $\Omega$, consistent with our numerical results, as shown in Fig.~\ref{fig_5}(a). In the intermediate regime of $\Omega$, the $J^\prime_x$ term dominates and a ferromagnetic order appears. For small $\Omega$, the effective model reduces to Eq.~(\ref{eff_soc_only}), and the spin vortex pattern dominates.
%Interesting, in $y$ direction, even though SLM coupling doesn't work, there is also a spiral order. The results are from the combined impact of SLM and SOAM, inducing the two-dimension vortex to become one-dimension spiral order.

In a realistic system, one can tune the balance of the two-spin components, which actually acts as an effective magnetic field $m_z$. Here, we can control the chemical potential difference of the two components to study the effect of the magnetic field. In Fig.~\ref{fig_6}, we fix the depth of optical lattice $V=9\, E_R$ with hopping $t\approx 0.045$, and study the many-body phase diagram as a function of the effective magnetic field $m_z$ and Raman-induced spin-flip hopping $\Omega$. When the effective magnetic field is large and negative, the spin-$\downarrow$ component supports a vortex structure, indicated by the local phase of the superfluid order parameter, as shown in inset of Fig.~\ref{fig_6}, or vice versa. For large enough Raman coupling, the two components are mixed, and the system supports a stripe pattern in the $x$ direction. We remake here that the phase diagram is similar to the one achieved by the SOAM experiments in continuous space~\cite{PhysRevLett.122.110402}, where the difference is vortex-antivortex pair phase for larger Raman coupling, instead of stripe order~\cite{PhysRevResearch.2.033152,PhysRevA.102.013316}, since we essentially include a Raman lattice in the $x$ direction.
%It clear to see the adjustment promotes the phase boundary of $\rm SF_{\uppercase\expandafter{\romannumeral2}}$ and $\rm SF_{\uppercase\expandafter{\romannumeral3}}$. It's the same as the results of SOAM coupling~\cite{PhysRevLett.122.110402,PhysRevA.102.013316}.
%detuning is unavoidable which would lift the degeneracy and hence influence the boundary of phase picture.
\\

{\section{conclusion and discussion}
In summary, we propose a scheme to investigate spin-orbital-angular-momentum coupling in strongly interacting bosonic gases in a two-dimensional square lattice. Using real-space dynamical mean-field theory, we obtain various quantum phases, including spin-vortex defect, composite vortex, canted-antiferromagnetic, and ferromagnetic insulating phases. Based on effective exchange models, we find that the spin-vortex texture is a result of the Dzyaloshinskii-Moriya interaction, induced by spin-orbital-angular-momentum coupling. Due to the competition of Dzyaloshinskii-Moriya and Heisenberg exchange interactions, various spin textures develop. In the superfluid, we find three quantum phases with conventional, stripe and vortex orders, characterized by the local phase of superfluid order parameters. Our study would be helpful to identify interesting many-body phases in future experiments. %Then we discuss the competition between SOAM and SLM with vanishing detuning, and find spiral phase, ACF and stripe phase. Exploring the influence of detuning, it only lifts the phase boundary. These findings would be helpful to identify the interesting phase in future experiments.

\section{acknowledgments}
We acknowledge helpful discussions with Kaijun Jiang and Keji Chen. This work is supported by the National Natural Science Foundation of China under Grants No.12074431, 11774428, and 11974423, and National Key RD program, Grant No. 2018YFA0306503. We acknowledge the Beijing Super Cloud Computing Center (BSCC) for providing HPC resources that have contributed to the research results reported within this paper.

\newpage
\begin{widetext}
\section{Appendix}
\renewcommand{\theequation}{S\arabic{equation}}
\renewcommand{\thefigure}{S\arabic{figure}}
\renewcommand{\bibnumfmt}[1]{[#1]}
\renewcommand{\citenumfont}[1]{#1}
\setcounter{equation}{0}
\setcounter{figure}{0}

\subsection{Single-particle Hamiltonian}
In our scheme, the Raman transition is $\Lambda$-type configuration with $\Omega_1(r)=\Omega_1 {\rm cos}(kx)$ along the $x$ direction and $\Omega_2(r)=\Omega_2e^{-2r^2/\rho^2}e^{-i2l\phi}$ along the $z$ direction. In the regime $\left| \Delta \right| \gg \Omega_{1,2}$, the single-photon transition between the ground and excited states is suppressed. We can adiabatically remove the excited state, and the system is effectively regarded as two-ground-state mixtures coupled by two-photon Raman processes. Including the two-photon Raman processes, we can obtain an effective spin-$1/2$ Hamiltonian
\begin{eqnarray}
H=\left( \begin{matrix} \frac{p^2}{2m}+\frac{\delta}{2}& \Omega^{\prime}\left( r \right) \cos kx\cdot e^{-i2l\phi}\\ \Omega^{\prime}\left( r \right) \cos kx\cdot e^{i2l\phi}& \frac{p^2}{2m}-\frac{\delta}{2}\\\end{matrix} \right),
\end{eqnarray}
where $\delta$ is a two-photon detuning, and $\Omega^{\prime}(r)=\Omega_1\Omega_2e^{-2r^2/\rho^2}/\Delta$ denotes the Raman Rabi frequency. After introducing the unity transformation to the single-particle wave function
\begin{eqnarray}
U&=&\left( \begin{matrix} e^{-il\phi}& 0\\ 0& e^{il\phi}\\\end{matrix} \right),
\end{eqnarray}
and Pauli matrix $\sigma$, we finally obtain
\begin{eqnarray}
H&=&-\frac{\hbar ^2}{2mr}\frac{\partial}{\partial _r}\left( r\frac{\partial}{\partial _r} \right) +\frac{\delta}{2}\sigma _z+\frac{\left( L_z-l\hbar \sigma _z \right)}{2mr^2}^2+\Omega ^{\prime}\left( r \right) \cos kx\cdot \sigma _x, \nonumber \\
\end{eqnarray}
where $L_z=-i\hbar \partial_z$ is the orbital angular momentum along the $z$ axis. Normally, the plan-wave laser is a LG beam with the intensity being suppressed near the trap center, which can influence experimental observations. Here, we instead consider a Gaussian-type Raman beam with orbital angular momentum, where such a Gaussian beam can be obtained by a quarter-wave plate~\cite{PhysRevLett.125.260407,Rafayelyan:17}. The waist of the plane-wave laser is set to $\rho=20$.

\subsection{Orbital angular momentum in the Wannier basis}

The angular momentum $t_{ij}$ is given by
\begin{eqnarray}
\label{Lz1}
\mathcal{H}_{L_z}&=&\int dx \Psi^{\dagger}(x)\frac{l\hbar}{mr^2}L_z \Psi(x) \nonumber \\
&=&\frac{l\hbar}{m}\int dx \Psi^{\dagger}(x)\frac{1}{r^2}L_z \Psi(x).
\end{eqnarray}
For a sufficient deep lattice, the field operator $\Psi(x)$ can be expanded in the lowest-band Wannier basis $\omega\left( \boldsymbol{x}-\boldsymbol{R}_i \right)$. Eq.~(\ref{Lz1}) can be rewritten as
\begin{eqnarray}
\mathcal{H}_{L_z}&=&\frac{l\hbar}{m}\sum_{\langle i,j\rangle} c^{\dagger}_i c_j\int{dx^3}\omega^{\ast}\left( \boldsymbol{x}-\boldsymbol{R}_i \right) \frac{1}{r^2}L_z \omega\left( \boldsymbol{x}-\boldsymbol{R}_j \right).
\end{eqnarray}
For nearest neighbors $i$ and $j$, we have the relation that
\begin{eqnarray}
\frac{1}{r_1^2}K_{i,j}<\int{dx^3}\omega^{\ast}\left( \boldsymbol{x}-\boldsymbol{R}_i \right) \frac{1}{r^2}L_z \omega\left( \boldsymbol{x}-\boldsymbol{R}_j \right)<\frac{1}{r_2^2}K_{i,j},\nonumber \\
\end{eqnarray}
where $r_1= {\rm max} (r_i,r_j)$, and $r_2={\rm min}(r_i,r_j)$, with $r_i$ ($r_j$) being the distance between site $i$ ($j$) and the trap center. For simplicity, we take the distance between the midpoint of sites $i$, $j$, and the trap center as the approximation of $r$, and denote it by $r^\prime$. Eq.~(\ref{Lz1}) reads
\begin{eqnarray}
\mathcal{H}_{L_z}&\approx&\frac{l\hbar}{mr^{\prime 2}}\sum_{\langle i,j\rangle} c^{\dagger}_i c_j K_{i,j},
\end{eqnarray}
with $K_{i,j}=\int{dx^3}\omega^{\ast}\left( \boldsymbol{x}-\boldsymbol{R}_i \right)L_z \omega\left( \boldsymbol{x}-\boldsymbol{R}_j \right)=-i\hbar\int{dx^3}\omega^{\ast}\left( \boldsymbol{x}-\boldsymbol{R}_i \right)(x\partial_y-y\partial_x) \omega\left( \boldsymbol{x}-\boldsymbol{R}_j \right)$. Generally, the Wannier function can be factorized into the $x$- and $y$-dependent parts for the deep lattice, and we finally obtain

\begin{figure}
\includegraphics[trim = 0mm 0mm 0mm 0mm, clip=true, width=0.5\textwidth]{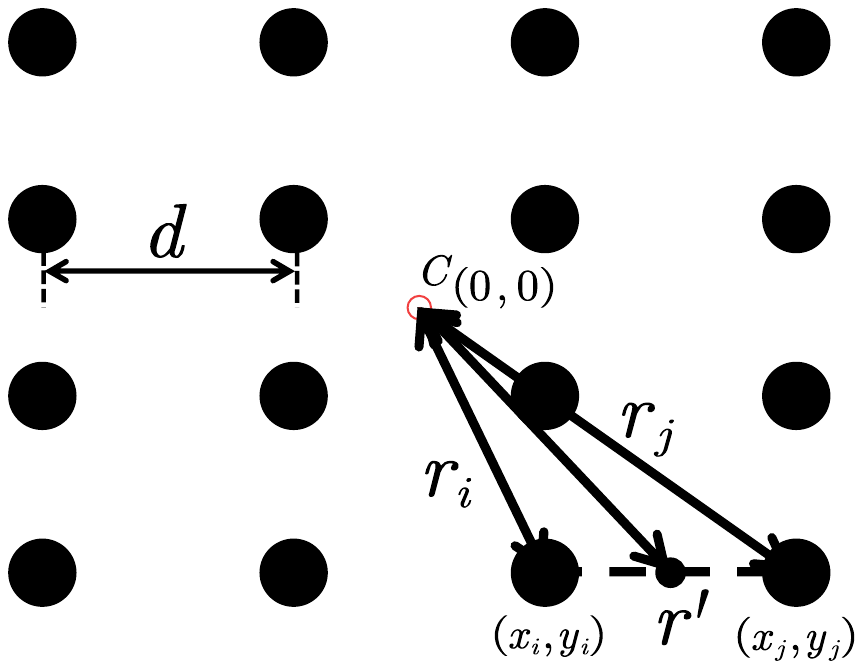}
\caption{(Color online) Schematic for a $4 \times 4$ lattice. $C$ is the trap center, and $r_i$ ($r_j$) is the distance between site $i$ ($j$) and the trap center $C$, with $d$ being the lattice constant. $r^{\prime}$ is the distance between the midpoint of sites $i$, $j$, and the trap center $C$. }
\label{fig_rotation}
\end{figure}

\begin{eqnarray}
K_{i,j}&=& -i\hbar \left( \int dx \omega^{\ast}(x-x_i)x \omega(x-x_j)\int dy \omega^{\ast}(y-y_i)\partial_y \omega(y-y_j)\right. \nonumber \\
&-&\left. \int dx \omega^{\ast}(x-x_i)\partial_x \omega(x-x_j)\int dy \omega^{\ast}(y-y_i)y \omega(y-y_j) \right).
\end{eqnarray}
For the discrete lattice system, $K_{ij}$ can be written in a product form
\begin{eqnarray}
K_{i,j}&=&-i\hbar \left(\frac{x_iy_j-x_jy_i}{d}\alpha \right),
\end{eqnarray}
with $\alpha=\int dx \omega^{\ast}\left(x-d \right)\partial_x \omega \left(x \right)$ and $d$ being the lattice constant, as shown in Fig.~\ref{fig_rotation}. We finally obtain the orbital angular momentum in the Wannier basis
\begin{eqnarray}
\mathcal{H}_{L_z}&=&\sum_{\langle i,j\rangle}-i\frac{l\hbar^2}{mr^{\prime 2}} \frac{x_iy_j-x_jy_i}{d} \left(c^{\dagger}_i c_j-c_i c^{\dagger}_j \right) \int dx \omega^{\ast}\left(x-d \right)\partial_x \omega \left(x \right)\nonumber \\
&=&\sum_{\langle i,j\rangle}i\frac{x_iy_j-x_jy_i}{r^{\prime 2}}\left(-\frac{l\hbar^2}{dm}\int dx \omega^{\ast}\left(x-d \right)\partial_x \omega \left(x \right)\right) \left(c^{\dagger}_i c_j-c_i c^{\dagger}_j \right).
\end{eqnarray}

\subsection{Effective exchange model}
The system can be described by an effective exchange model in the deep Mott-insulating regime. To derive the effective model, we first divide the Hilbert space according into site occupations for filling $n=1$. We define the operators $\mathcal{P}$ and $\mathcal{Q}$ , which denote the projection into the Mott state subspace $\mathcal{H}_P$ and the perpendicular subspace $\mathcal{H}_Q$. For a Hamiltonian $H$, the Schr{\"o}dinger equation reads
\begin{eqnarray}
H\left|\psi \right>=E\left|\psi\right>.
\end{eqnarray}
With the unity operator $1=\mathcal{P}+\mathcal{Q}$, we obtain
\begin{eqnarray}
\label{eq:1}
H(\mathcal{P}+\mathcal{Q})\left|\psi\right>=E(\mathcal{P}+\mathcal{Q})\left|\psi\right>.
\end{eqnarray}
Multiplying by $\mathcal{P}$ and $\mathcal{Q}$ the left side of Eq.~(\ref{eq:1}) results in
\begin{eqnarray}
\label{eq:2}
(\mathcal{P}H\mathcal{P}+\mathcal{P}H\mathcal{Q})|\psi\rangle&=&E\mathcal{P}|\psi\rangle, \\
\label{eq:3}
(\mathcal{Q}H\mathcal{P}+\mathcal{Q}H\mathcal{Q})|\psi\rangle&=&E\mathcal{Q}|\psi\rangle.
\end{eqnarray}
Eq.~(\ref{eq:3}) can be rewritten with the projection operator relation $\mathcal{Q}^2=\mathcal{Q}$,
\begin{eqnarray}
\label{eq:4}
\mathcal{Q}\left|\psi\right>=\frac{1}{E-\mathcal{Q}H\mathcal{Q}}\mathcal{Q}H\mathcal{P}|\psi\rangle.
\end{eqnarray}
Inserting~(\ref{eq:4}) into~(\ref{eq:2}), we obtain an equation for $\mathcal{P}|\psi\rangle$,
\begin{eqnarray}
\label{eq:5}
(\mathcal{P}H\mathcal{Q}\frac{1}{E-\mathcal{Q}H\mathcal{Q}}\mathcal{Q}H\mathcal{P})\mathcal{P}|\psi\rangle=E\mathcal{P}|\psi\rangle.
\end{eqnarray}

For Hamiltonian $H$, we can divide it into two parts $H=H_t+H_U$, where $H_t$ and $H_U$ are the hopping and interaction parts of $H$, respectively. Eq.~(\ref{eq:5}) can be rewritten as
\begin{eqnarray}
\label{eq:6}
(\mathcal{P}H_t\mathcal{Q}\frac{1}{E-\mathcal{Q}H_U\mathcal{Q}-\mathcal{Q}H_t\mathcal{Q}}\mathcal{Q}H_t\mathcal{P})\mathcal{P}|\psi\rangle=E\mathcal{P}|\psi\rangle,
\end{eqnarray}
with $\mathcal{P}H_U\mathcal{P}$, $\mathcal{P}H_U\mathcal{Q}$ and $\mathcal{P}H_t\mathcal{P}$ being zero. Because $E\sim\frac{t^2}{U}\sim0$ in the Mott phase, we take $\frac{1}{E-\mathcal{Q}H_U\mathcal{Q}-\mathcal{Q}H_t\mathcal{Q}}\approx\frac{1}{-\mathcal{Q}H_U\mathcal{Q}-\mathcal{Q}H_t\mathcal{Q}}$. %The expansion of $\frac{1}{-\mathcal{Q}H_U\mathcal{Q}-\mathcal{Q}H_t\mathcal{Q}}$ by
Using the expansion $\frac{1}{A-B}=\frac{1}{A}\Sigma^{\infty}_{n=0}(B\frac{1}{A})^n$, with $A=-\mathcal{Q}H_U\mathcal{Q}$ and $B=\mathcal{Q}H_t\mathcal{Q}$, we finally obtain
\begin{eqnarray}
\label{eq:eff_ham}
\mathcal{H}_{eff}&=&\mathcal{P}H_t\mathcal{Q}\frac{1}{-\mathcal{Q}H_U\mathcal{Q}}\cdot \Sigma^{\infty}_{n=0}(\mathcal{Q}H_t\mathcal{Q}\frac{1}{\mathcal{Q}H_U\mathcal{Q}})^n \mathcal{Q}H_t\mathcal{P}.
\end{eqnarray}
Normally, we only need to include nearest-neighbor terms in the effective Hamiltonian with $n=0$, i.e.
\begin{eqnarray}
\mathcal{H}_{eff}&=&\mathcal{P}H_t\mathcal{Q}\frac{1}{-\mathcal{Q}H_U\mathcal{Q}}\mathcal{Q}H_t\mathcal{P}.
\label{effective}
\end{eqnarray}
%Eq.~(\ref{effective}) is the starting point in the derivation of effective Hamiltonian.
In the tight-binding regime, the subspace $\mathcal{H}_{P}$ of the states with half filling for a two-site problem is
\begin{eqnarray}
\mathcal{H}_{P}: \left\{ \left|\uparrow,\uparrow \right> , \left| \uparrow, \downarrow \right> , \left| \downarrow ,\downarrow \right> , \left| \downarrow ,\downarrow \right> \right\},
\end{eqnarray}
where $\left| \sigma, \sigma^\prime \right>$ denotes the spin state $\sigma$ in the left site and $\sigma^\prime$ in the right one. The subspace $\mathcal{H}_{P}$ for doubly occupied sites is
\begin{eqnarray}
\mathcal{H}_{Q}: \left\{ \left| \uparrow \uparrow\right> , \left| \uparrow \downarrow \right> , \left| \downarrow \downarrow \right> \right\}.
\end{eqnarray}
In a matrix form, $\left( \mathcal{Q} H_U\mathcal{Q} \right) ^{-1}$ is given by
\begin{eqnarray}
\left( \mathcal{Q} H_U\mathcal{Q} \right) ^{-1}&=&\left( \begin{matrix} \frac{1}{U}& 0& 0\\ 0& \frac{1}{U}& 0\\ 0& 0& \frac{1}{U_{\uparrow \downarrow}}\\\end{matrix} \right).
\end{eqnarray}
According to Eq.~(\ref{effective}), we obtain the final effective Hamiltonian
\begin{eqnarray}
\mathcal{H}_{eff}&=&-\left( \frac{4\left( t^2+t_{ij}^{2} \right)}{U}+\frac{2\Omega ^2}{U_{\uparrow \downarrow}} \right) \left( n_{i,\uparrow}n_{j,\uparrow}+n_{i,\downarrow}n_{j,\downarrow} \right) -\left( \frac{2\left( t^2+t_{ij}^{2} \right)}{U_{\uparrow \downarrow}}+\frac{4\Omega ^2}{U} \right) \left( n_{i,\uparrow}n_{j,\downarrow}+n_{i,\downarrow}n_{j,\uparrow} \right) \nonumber\\
&-&\frac{2\left( -t+it_{ij} \right) ^2}{U_{\uparrow \downarrow}}c_{i,\downarrow}^{\dagger}c_{i,\uparrow}c_{j,\uparrow}^{\dagger}c_{j,\downarrow}-\frac{2\left( -t-it_{ij} \right) ^2}{U_{\uparrow \downarrow}}c_{i,\uparrow}^{\dagger}c_{i,\downarrow}c_{j,\downarrow}^{\dagger}c_{j,\uparrow}\nonumber \\
&-&\frac{2\Omega ^2}{U_{\uparrow \downarrow}}\left( c_{i,\uparrow}^{\dagger}c_{i,\downarrow}c_{j,\uparrow}^{\dagger}c_{j,\downarrow}+c_{i,\downarrow}^{\dagger}c_{i,\uparrow}c_{j,\downarrow}^{\dagger}c_{j,\uparrow} \right)
-(-1)^{i_x}\left( \frac{2\Omega \left( -t-it_{ij} \right)}{U}+\frac{2\Omega \left( -t-it_{ij} \right)}{U_{\uparrow \downarrow}} \right) \left( c_{j,\downarrow}^{\dagger}c_{j,\uparrow}+c_{i,\uparrow}^{\dagger}c_{i,\downarrow} \right) \nonumber \\
&-&(-1)^{i_x}\left( \frac{2\Omega \left( -t+it_{ij} \right)}{U_{\uparrow \downarrow}}+\frac{2\Omega \left( -t+it_{ij} \right)}{U} \right) \left( c_{j,\uparrow}^{\dagger}c_{j,\downarrow}+c_{i,\downarrow}^{\dagger}c_{i,\uparrow} \right). \nonumber \\
\label{effec1}
\end{eqnarray}
Introducing the pseudospin operator as follows
\begin{eqnarray}
S^x_i&=&\frac{1}{2} \left( c^{\dagger}_{i,\uparrow}c_{i,\downarrow} + c^{\dagger}_{i,\downarrow} c_{i,\uparrow} \right) \\
S^y_i&=&\frac{1}{2i} \left( c^{\dagger}_{i,\uparrow}c_{i,\downarrow} - c^{\dagger}_{i,\downarrow} c_{i,\uparrow} \right) \\
S^z_i&=&\frac{1}{2} \left( n_{i,\uparrow}- n_{i,\downarrow} \right),
\end{eqnarray}
Eq.~(\ref{effec1}) can be rewritten as
\begin{eqnarray}
\mathcal{H}_{eff}&=&-4\left( \frac{2\left( t^2+t_{ij}^{2} \right)}{U}+\frac{\Omega ^2}{U_{\uparrow \downarrow}}-\frac{\left( t^2+t_{ij}^{2} \right)}{U_{\uparrow \downarrow}}-\frac{2\Omega ^2}{U} \right) S_{i}^{z}S_{j}^{z} -4\left( \frac{\left( t^2-t_{ij}^{2} \right)}{U_{\uparrow \downarrow}}+\frac{\Omega ^2}{U_{\uparrow \downarrow}} \right) S_{i}^{x}S_{j}^{x}
\nonumber \\
&-&4\left( \frac{\left( t^2-t_{ij}^{2} \right)}{U_{\uparrow \downarrow}} -\frac{\Omega ^2}{U_{\uparrow \downarrow}} \right) S_{i}^{y}S_{j}^{y}
-\frac{8tt_{ij}}{U_{\uparrow \downarrow}}\left( S_{i}\times S_{j} \right)_z \nonumber \\
&+&(-1)^{i_x}\Omega t\left( \frac{4}{U}+\frac{4}{U_{\uparrow \downarrow}} \right) \left( S_{i}^{x}+S_{j}^{x} \right) -(-1)^{i_x}\Omega t_{ij}\left( \frac{4}{U}+\frac{4}{U_{\uparrow \downarrow}} \right) \left( S_{i}^{y}-S_{j}^{y} \right).
\end{eqnarray}
The result can be easily extended to the case with vanishing $\Omega=0$, and it reads
\begin{eqnarray}
\mathcal{H}_{eff}&=&-4\left( \frac{2\left( t^2+t_{ij}^{2} \right)}{U}-\frac{\left( t^2+t_{ij}^{2} \right)}{U_{\uparrow \downarrow}} \right) S_{i}^{z}S_{j}^{z} - \frac{4\left( t^2-t_{ij}^{2} \right)}{U_{\uparrow \downarrow}} S_{i}^{x}S_{j}^{x}
\nonumber \\
&-& \frac{4\left( t^2-t_{ij}^{2} \right)}{U_{\uparrow \downarrow}} S_{i}^{y}S_{j}^{y}
-\frac{8tt_{ij}}{U_{\uparrow \downarrow}}\left( S_{i}\times S_{j} \right)_z. \nonumber \\
\end{eqnarray}

\end{widetext}

\bibliography{references}

\begin{thebibliography}{77}
\bibitem[{bat()}]{bath}
\bibinfo{note}{In our simulations, the number of normal bath orbitals is
  $N_{\rm bath}=4$ with the maximum occupation number for each normal bath
  orbital $N_l=2$. The Fock space is truncated at a maximum occupation number
  $N_{\rm fock}=5$.}
\bibitem[{ban()}]{band_structure}
\bibinfo{note}{In the band-structure simulations, we consider a system of
  ultracold $^{87}$Rb atoms with s-wave scattering length $a^{\uparrow
  \downarrow}_s \approx 100\, a_{B}$ and $a^{\uparrow,\downarrow}_s
  /a^{\uparrow \downarrow}_s \approx 1.01$, with $a_B$ being the Bohr radius.
  To achieve a two-dimensional square lattice formed by standing-wave lasers
  with a wave length 738 nm, a strong confiment is assumed in the z direction
  with a depth $V_z=100E_R$}.
\end{thebibliography}

\end{document}